# Interface Condition for the Darcy Velocity at the Water-oil Flood Front in the Porous Medium

Xiaolong Peng, Yong Liu, Baosheng Liang
(State Key Laboratory of Oil and Gas Reservoir Geology and Exploitation, Southwest Petroleum University; Chengdu, 610500, China)

## ABSTRACT

Flood front is the jump interface where fluids distribute discontinuously, whose interface condition is the theoretical basis of a mathematical model of the multiphase flow in porous medium. The conventional interface condition at the jump interface is expressed as the continuous Darcy velocity and fluid pressure (named CVCM ). Our study has inspected this conclusions. First, it is revealed that the principle of mass conservation has no direct relation to the velocity conservation, and the former is not the true foundation of the later, because the former only reflects the kinetic characteristic of the fluid particles at one position(the interface), but not the neighborhood of the interface which required by the later. Then the reasonableness of CVCM is queried from the following three aspects:(1)Using Mukat's two phase seepage equation and the mathematical method of apagoge, we have disproved the continuity of each fluid velocity;(2)Since the analytical solution of the equation of Buckley-Leveret equations is acquirable, its velocity jumps at the flood front presents an appropriate example to disprove the CVCM;(3) The numerical simulation model gives impractical result that flood front would stop moving if CVCM were used to calculate the velocities at the interface between two gridcells. Subsequently, a new one, termed as Jump Velocity Condition Model (JVCM), is deduced from Muskat's two phase seepage equations and Darcy's law without taking account of the capillary force and compressibility of rocks and fluids. Finally, several cases are presented. And the comparisons of the velocity, pressure difference and the front position, which are given by JVCM, CVCM and SPU, have shown that the result of JVCM is the closest to the exact solution.

| Nomenclature | | | |
|---|---|---|---|
| $f$ | fractional flow, dimensionless | $t$ | time,T |
| $K$ | absolute permeabillity,$L^2$ | $x$ | distance,L |
| $k_c$ | relative peameability,dimensionless | $\Gamma$ | interface, symbol. |
| $\phi$ | porosity, dimensionless | $\lambda$ | mobility, $LM^{-1}T$ |
| $\mu$ | viscocity,$L^{-1}MT^{-1}$ | **Subscripts** | |
| $s$ | saturation, dimensionless; | $w$ | water |
| $S$ | saturation, dimensionless; | $n$ | non-wet fluid, or oil |
| $P$ | pressure, $L^{-1}MT^{-2}$ | $o$ | oil |
| $v$ | Darcy velocity,$ML^{-1}$ | $t$ | total |
| $A_r$ | cross section area;$L^2$ | $-$ | limit on the upstream side |
| | | $+$ | limit on the downstream side |
| | | in | inflow |
| | | out | outflow |

## 1. Introduction

There are various abrupt interfaces, which are also treated as discontinuous interfaces in reservoirs, such as sudden changes of rock properties, original oil-water interfaces, and flood fronts. The seepage problem of discontinuous interfaces is called the Neumann problem. Its interface conditions at the jump interface is called Jump Condition, usually including the coupling relationship of the Darcy velocity and fluid pressure on the different sides of the interface, which constitutes the pre-condition for a complete seepage differential mathematical model and reservoir simulation model.

There are a variety of conditions at the interface, such as condition for mass conservation, momentum conservation, and energy conservation, etc. But in terms of numerical simulation of multiphase flow in porous media, the key conditions are the condition for Darcy velocity and condition for pressures. Many very important technology about it are associated with the interface conditions, e.g. Designing FDM (Aziz. and Settari, 1979) or FVM (LeVeque, 2002) or FEM (Reddy Gartling 2010; Costa ,Oliveira, Baliga, et.al.2004) to numerically solve the seepage differential equations, domain decomposition and parallel computing (Nataf, 2002), creating the formula for the inter-porosity flow (Wu, Qin, 2009), measuring the fluid flux between the rocks and hydraulic fracture or between the reservoir and the wellbore, multiscale numerical simulation ( Lunati, Jenny, 2006), tracer testing analysis (Haggerty, McKenna, Meigs, 2000; Datta-Gupta, King, 1995).

The conventional theory of the multiphase flow in the porous media contends that in accordance with the mass conservation principle, each fluid velocity at flood front is conservational in each phase (Szymkiewicz, 2012; Valdes-Parada, Espinosa-Paredes, 2005; Brenner, Cances, Hilhorst, 2013; Szymkiewicz, 2012; Hassanizadeh, Gray, 1989; Doster, Hilfer, 2011; Ohlberger, Schweizer, 2007; Mozolevski, Schuh, 2013; Huber, Helmig, 2000). So:

$$v_\ell^- \cdot N_\Gamma = v_\ell^+ \cdot N_\Gamma \qquad (1)$$

Where $N_\Gamma$ is a unit vector normal to the interface $\Gamma$, and $v_\ell^- \triangleq v_\ell(x^-)$ $x^- = \lim_{\varepsilon \to 0, \varepsilon > 0}(x-\varepsilon)$, and $v_\ell^- \triangleq v_\ell(x^+)$ $x^+ = \lim_{\varepsilon \to 0, \varepsilon > 0}(x+\varepsilon)$. For the one-dimensional incompressible seepage question, the above formula can be simplified as

$$v_\ell^- = v_\ell^+ \qquad (2)$$

In addition, its second condition is that each fluid pressure at the flood front is also continuous, namely

$$P_\ell^- = P_\ell^+ \quad (\ell = o, w) \qquad (3)$$

Where $P_\ell^- \triangleq P_\ell(x^-)$, and $P_\ell^+ \triangleq P_\ell(x^+)$. For convenience of description, the (1) (3) or (2) (3) are collectively marked as Continuous Velocity Condition Model (CVCM).

However, we think the conventional interface conditions, CVCM, are not reasonable under the condition that fluids distributed discontinuously at the flood front. First, we can explain that mass conservation principle has no direct relation to the velocity conservation, and the former is not real strong foundation for the later. Then the reasonability of CVCM is queried from following three ways:① disproving the continuity of each fluid velocity in the way of mathematics;②via solving the equation of Buckley-Leveret equations, presenting a appropriate example of velocity jumps;③showing an impractical phenomenon that flood front would stop moving, calculated by the numerical simulated model in which CVCM is used to calculate the numerical flux at the interface between two gridcells.

Subsequently, a new one, termed as Jump Velocity Condition Model (JVCM), is deduced from Muscat's two phase seepage equations without taking account of the capillary force and compressibility of rocks and fluids.

## 2. Analysis of the Jump Condition on the Flood Front Interface

Supposing that the oil-water seepage flow in one-dimensional horizontal reservoir, ignoring the influence of compressibility, capillary force, fluid gravity, as flood front interface is discontinuous, the reservoir can be divided into 2 continuous sub-domains, namely $\Omega_1$ and $\Omega_2$, where the Muskat two-phase seepage differential equations can be established respectively (Muskat,1937):

$$\begin{cases} \nabla \cdot v_o + \dfrac{\partial(\phi S_o)}{dt} = 0 \\ \nabla \cdot v_w + \dfrac{\partial(\phi S_w)}{dt} = 0 \end{cases} \quad \{(x,t) \mid x \in (\bar{\Omega}_1 \cup \bar{\Omega}_2) \setminus \Gamma, t \geq 0; \Gamma = \bar{\Omega}_1 \cap \bar{\Omega}_2\} \tag{4}$$

Motion equation

$$v_\ell = -K\lambda_\ell \nabla P_\ell \tag{5}$$

Where $\lambda_\ell \triangleq \dfrac{k_{r\ell}}{\mu_\ell}$. Define the fractional flow coefficient of phase $\ell$.

$$f_\ell \triangleq \dfrac{\lambda_\ell}{\lambda_o + \lambda_w} \tag{6}$$

According to the seepage mechanics theory, fractional flow coefficient is a continuous function of saturation $f_\ell \triangleq f_\ell(S_\ell)$, generally meeting the monotone condition. Then equation (7) can be rewritten as

$$v_\ell = -f_\ell v_t \tag{7}$$

Add the two equations of Equation (4) together, takes the equation

$$\dfrac{\partial(v_o + v_w)}{\partial x} = 0 \text{ Or } v_o + v_w \text{ is constant.} \tag{8}$$

With defining $v_t \triangleq v_o + v_w$ and $\lambda_t \triangleq \lambda_o + \lambda_w$, and applying Equ.(5), results in:

$$v_t = \lambda_t K \dfrac{\partial P}{\partial x} \tag{9}$$

Both sides of Equation.(9) divided by $\lambda_t K$ and do the finite integral in the domain of $(x_1, x_2)$, Equ.(9) can be altered into:

$$v_t = -\dfrac{P_1 - P_2}{\int_{x_1}^{x_\Gamma} \dfrac{dx}{\lambda_t K} + \int_{x_\Gamma}^{x_2} \dfrac{dx}{\lambda_t K}} \tag{10}$$

Where $P_1 \triangleq P(x_1)$ and $P_2 \triangleq P(x_2)$.

The following shows how proofs by contradiction is employed to prove $v_\ell^- \neq v_\ell^+ (\ell = o, w)$ on the displacement interface $\Gamma$.

## 2.1 Relationship between Mass Conventional Principle and Interface Condition for Darcy Velocities

As shown in Figure 1, suppose there is a particle group M at flood front $\Gamma$, whose coordinate is x and velocity is $v$. There are another two particle groups adjacent to $M$, viz, $M^+$ and $M^+$, whose coordinate are $x^-$ and $x^+$ respectively. Certainly the $x^-$ and $x^+$ adjoin x from different direction.

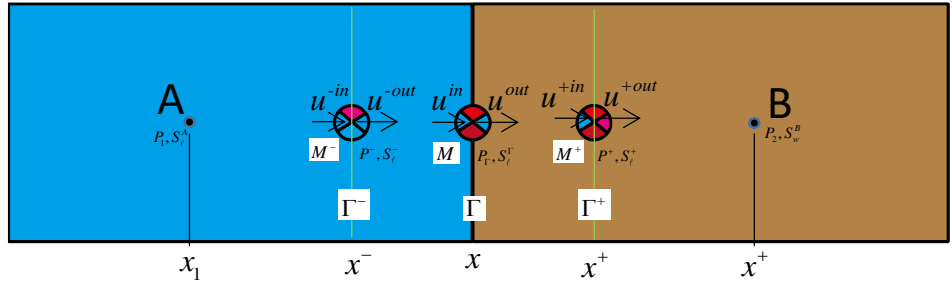

Fig.1 Schemes for the fluid flow at the jump interface

(1)The physicals tells us that the nature of motion is to transport the particle from one place to another place. In Fig.1, at x, M is flowing through the interface $\Gamma$ with velocity $v$. That is to say, M is flowing into the interface $\Gamma$ whose inflow rate indicates by $u_\ell^{in}$, and simultaneously M is also flowing out of $\Gamma$ whose outflow rate indicates by $u_\ell^{out}$. According to the mass conservation principle, at any time the inflow and outflow at the interface $\Gamma$ must be equal, therefore:

$$u_\ell^{in}(x) = u_\ell^{out}(x) \qquad (11)$$

(2)Although $u_\ell^{in}$ and $u_\ell^{out}$ have the same value, the same direction as $v_\ell$, they are just the attributes of the velocities $v_\ell (\ell = w, o)$ at $\Gamma$. Since $u_\ell^{in}$ cannot be regarded as $v_\ell^-$ (Darcy velocity of fluid $\ell$ at $\Gamma^-$) and $u_\ell^{out}$ regard cannot be treated as $v_\ell^+$ (Darcy velocity of fluid $\ell$ at $\Gamma^+$), we cannot use the equation Equ.11 to make the conclusion $v_\ell^- = v_\ell^+$. However, the CVCM claimed the interface condition for Darcy velocity is $v_\ell^- = v_\ell^+$, and its alleged foundation is mass conservation equation principle, which implies the principle of mass conservation is not the real foundation of the CVCM, and if Darcy velocities are not conservational does not mean they violates the principle of mass conservation.

(3) Although there are two term s in **Equ.2**, the both reflect the motion of one particle group ($M$). However, **Equ.6** has points to two different particle groups ($M^-$ and $M^+$) which locates at different positions($\Gamma^-$ and $\Gamma^+$). Thus can further explains that principle mass conservation has no bound relation to the CVCM.

(4) While M is moving from its original position, $M^-$ is going to fill the empty space left by $M$. Under the precondition of saturated multiphase flow, namely the pore space is always filled with fluid, without considering the compressibility of fluid and porous medium, the volume of $M^-$ and that $M^+$ should be equal, but it is not strictly required that $M^-$ and $M^+$ are the same kind of fluid. So we can have $v_o^- + v_w^- = v_o + v_w$ (as same as Equation. (9)), but not ensure $v_o^- = v_o^-$ or $v_o^+ = v_o^+$.

(5) Actually, according to the principle of the mass conversation, only when $v_w^- > v_w^+$, The oil, located between $\Gamma^-$ and $\Gamma^+$, can be replaced by the water.

## 2.2 Prove the Inequality between the Darcy Velocities on the Both Sides of Flood Front via Apagoge

**Theorem 1:** in regard to incompressible oil-water flow in the one dimensional horizontal reservoir (seen in Equ.**3**), fractional flow coefficient $f_\ell$ is a monotone function of saturation $S_\ell$. If both the flood front saturation (marked as $S_{\ell f}$) and the initial saturation ($S_{\ell i}$, $S_{\ell i} \neq S_{\ell f}$) are known, the Darcy velocity of each fluid is not conservational.

**Proof by Contradiction:**

[A1] Assume the Darcy velocity is conservational at the flood front. So it is continuous in the whole reservoir. Based on the mathematical notion of continuity, there must exist the following equations within any infinitesimal neighborhood of $x_\Gamma$:

$$\lim_{\substack{x^- < x_\Gamma \\ x^- \to x_\Gamma}} v_\ell(x^-) = \lim_{\substack{x^+ > x_\Gamma \\ x^+ \to x_\Gamma}} v_\ell(x^+) = v_\ell(x_\Gamma) \tag{12}$$

Or

$$\lim_{(x^- - x^+) \to 0} \left| \lim_{\substack{x^- < x_\Gamma \\ x^- \to x_\Gamma}} v_\ell(x^-) - \lim_{\substack{x^+ > x_\Gamma \\ x^+ \to x_\Gamma}} v_\ell(x^+) \right| = 0$$

On the other hand, adding the two equations in Equ.3 together will result in $\nabla \cdot (v_o + v_w) = 0$. Apparently, $v_t \stackrel{\triangle}{=} v_o + v_w$ is a constant. It can be concluded by Formula (4) that

$$v_\ell = f_\ell(S_\ell) v_t \tag{13}$$

Therefore,

$$\lim_{\substack{x^- < x_\Gamma \\ x^- \to x_\Gamma}} v_\ell(x^-) = v_t \lim_{\substack{x^- < x_\Gamma \\ x^- \to x_\Gamma}} f_\ell(S_\ell(x^-)) = v_t f_\ell(S_{\ell f}) \tag{14}$$

$$\lim_{\substack{x^+ > x_\Gamma \\ x^+ \to x_\Gamma}} v_\ell(x^+) = v_t \lim_{\substack{x^+ > x_\Gamma \\ x^+ \to x_\Gamma}} f_\ell(S_\ell(x^+)) = v_t f_\ell(S_{\ell i}) \tag{15}$$

As the flood front is discontinuous interface of fluid distribution, namely, $S_w^- = S_{wf} \neq S_w^+ = S_{wi}$, so

$$\lim_{\substack{x^- < x_\Gamma \\ x^- \to x_\Gamma}} v_\ell(x^-) \neq \lim_{\substack{x^+ > x_\Gamma \\ x^+ \to x_\Gamma}} v_\ell(x^+) \tag{16}$$

Contradiction can be found between Formula (16) and (12), which states that the assumption [A1] is untenable and the Jump Condition expressed in equation (1)- (2) is not reasonable. So the Theorem.1 gets proved.

## 2.3 Example 1: Velocity Jump Proposed by Buckley-Leveret Model.

The analysis solution of the equation of Buckley-Leveret equations is acquirable (Doster, Hilfer, 2011; Buckley, Leverett, 1942). So its velocity field is a strong evidence to prove or disprove the CVCM.

For example, similar to Figure.1, there is a horizontal linear reservoir with 100 length, and 1m² cross section area. The oil-water relative permeability curve has been shown in Fig.2-(a). The oil viscosity $\mu_o$ is 2cp, and that of water $\mu_w$ is 0.3cp. At the left end, the water saturation is 1 during all the time of $t \geq 0$, where water has been injected into the reservoir with the fixed rate 1m³/day. The initial water saturation is 0. When $t=0$, all the reservoir is filled with oil, that is, $S_o(x,t=0)=1, x \in (0,100]$.

First, according to relative permeability and viscosities of oil and water, the front water ration can be cumulated, we get $S_{wf}$ =0.407. Then the water saturation at any position in the reservoir can be obtained through solving the Buckley-Leveret model equation, which is displayed in figure 2-(b). Finally, according to Equation (5), the fluid velocities of water or oil at any position are calculated, as shown in the Fig. 2-(c), in which the plot of $v_w - x$ shows the velocity of water is discontinuous rather than continuous at the flood front.

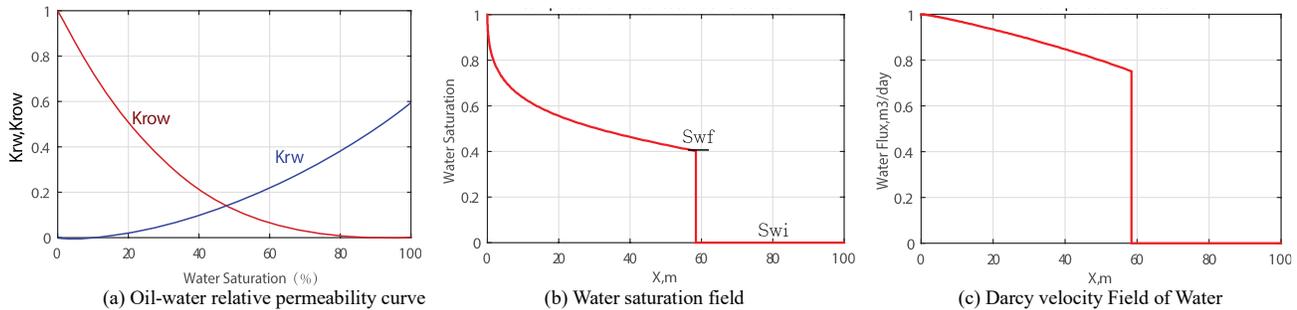

(a) Oil-water relative permeability curve  (b) Water saturation field  (c) Darcy velocity Field of Water

Fig.2 An example of water saturation field and Darcy velocity field of Buckley-Leveret impressible two-phase flow; (a) relative permeability curve; (b) water saturation field; (c) water Darcy-velocity field.

## 2.3 Example 2: CVCM cause the Impractical Virtual Phenomenon that Flood Stops Moving in the Numerical Simulation Model

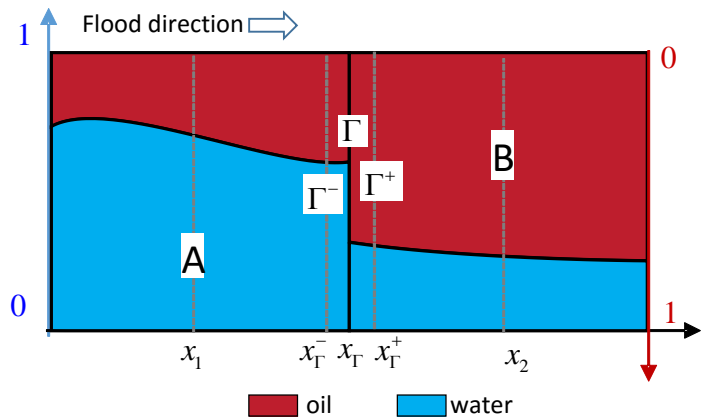

Fig.3 Scheme for the water-oil displacement at given time $t^n$

In this section, we present another example to disprove CVCM .that is ,the numerical simulated model gives a unreasonable result that flood front would stop moving if CVCM is used to calculate the numerical flux at the interface between two gridcells.

Firstly, in Figure.3, there are two grid cells, Cell A and Cell B. $x_A$ and $x_B$ represent their center position respectively. At the time $t = t^n$ the interface of the two grid cells located at the flood front ($\Gamma$), and the water situation of cell A is $S_w^A > S_{wi}^A$, that of the Cell B $S_w^B = S_{wi}^B$.

According to the theory of finite difference method (FDM; Aziz, Settari, 1979), the multiphase Darcy law can be expressed approximately as below:

$$v_\ell^A = \frac{\lambda_\ell^A K^A}{(x_\Gamma - x_A)}(P_A - P_\Gamma), \quad v_\ell^B = \frac{\lambda_\ell^B K^B}{(x_B - x_\Gamma)}(P_\Gamma - P_B) \tag{17}$$

Employing Equation(17) and CVCM can obtains the numerical flux at the interface $\Gamma$

$$q_\ell^\Gamma = \frac{\dfrac{\lambda_\ell^A K^A}{(x_\Gamma - x_A)}\dfrac{\lambda_\ell^B K^B}{(x_B - x_\Gamma)}}{\dfrac{\lambda_\ell^A K^A}{(x_\Gamma - x_A)} + \dfrac{\lambda_\ell^B K^B}{(x_B - x_\Gamma)}}(P_A - P_B) \quad (\ell = o, w) \tag{18}$$

At the time $t = t^n$, because $S_w^B = S_{wi}^B$, then $k_{rw}(s_{wi}^B) = 0$, and $\lambda_w^B \triangleq \lambda_w(s_{wi}^B) = 0$, certainly,

$$q_w^{\Gamma,n} \triangleq q_w(t = t^n, x = x_\Gamma) = 0 \tag{19}$$

Because it is known from the equation (19), the water flow rate is zero, accordingly in the time interval $t^n \sim t^{n+1}$, $\int_{t^n}^{t^{n+1}} q_w^\Gamma dt = 0$. The physical interpretation goes that in this period, the water inflow into the Cell B is 0 and the flood front stops moving. The same procedure may be easily adapted to obtain $\int_{t^{n+1}}^{t^{n+2}} v_w^\Gamma dt = 0$ in the time interval of $t^{n+1} \sim t^{n+2}$. By that analysis, it can be concluded that the water in formation can never go through $\Gamma$, which is obviously opposed to the physical truth and is exactly the reason why Harmonic Average is not adopted for the linearized processing of relative permeability during setting up the numerical model for reservoir numerical simulation. This case furthermore explains the irrationality of the CVCM.

On the other hand, if people confess the velocity field is discontinuous at the interface $\Gamma$, such problem can be avoided. At the moment of $t^n$, the water at $x^+$ will run through $\Gamma^+$ at the speed of $v_w^+$, and at $x^-$ run through $\Gamma^-$ at the speed $v^-$. Due to the inequality of velocity $v_w^-$ and $v_w^+$, $\Gamma^-$ - $\Gamma^+$ interval accumulates more water and less oil, which is consistent with displacement. Otherwise, like the results given by CVCM, the water saturation in $\Gamma^-$ - $\Gamma^+$ interval won't change, which mismatches the displacement.

## 3. The New Model for the Jump Conditions at the Flood Front

### 3.1 Establish the Interface Condition for Mass

If the fluid saturation field is known, using equation (5), the Interface for the mass can be obtained:

$$v_\ell^- - v_\ell^+ = (f_\ell^- - f_\ell^+)v_t \tag{20}$$

However, for better illustration, we are going to derive the interface condition for mass from the perspective of the process of water-oil displacement.

From micro perspective, the flowing space of fluid is a series of intricate capillary network. The one-dimensional issue can be simplified as capillary buddle, as shown in Fig.4 (1) ~ (2). If it is the case that oil-water flow at the state of continuous phases, then oil and water are always trying to occupy different flow path respectively. The higher the fluid saturation is, the lower the velocity, and the more channels occupied, whereas, the less flow path occupied. Then it is inevitable that discontinuous interface would occur where (as shown in Fig.5) one fluid displaces the other. This way of displacing is referred to as different fluid displacing by the Literature (Peng, Du, Liang, et.al 2007); In the other flow path, fluid of same kind displaces each other, which is addressed as the same fluid displacing, like water-displacing-water or oil-displacing-oil.

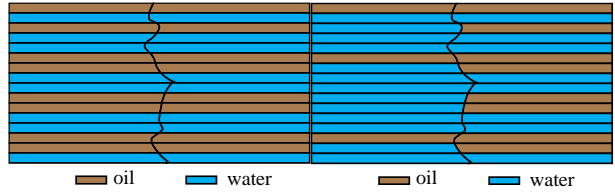

(1) Continuous interface  (2) Discontinuous interface
Fig.4 Scheme for fluid distribution beside the continuous interface and discontinuous interface

Mark the flux produced in water-displacing-water in Fig.2 as q1, that in oil-displacing-oil as q2 (as shown in Fig.5), and that in water-displacing-oil as q3. In the discontinuous interface, as compressibility is ignored, the volume of displacing phase flown in any seepage path equates to that of the outflow. In addition, without the influence of capillary force and gravity, the pressure in discontinuous interface is equivalent, as shown in the following:

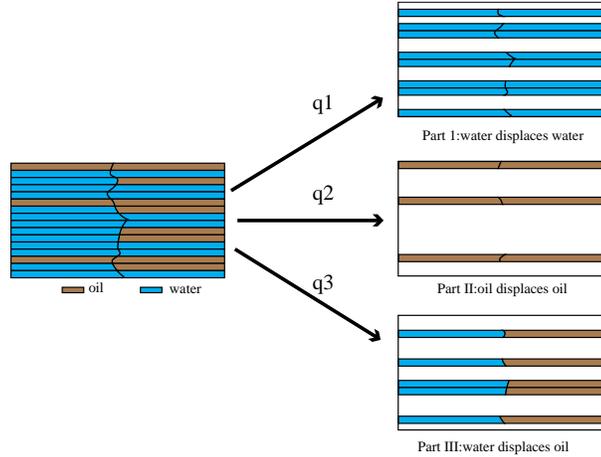

Fig.5 Flux decomposition and micro infinitesimal partition beside the discontinuous interface.

(1) Inside the Infinitesimal A, all oil involves in-phase displacement only (as shown in Fig.5 Part II). Therefore,

$$\bar{v}_o = \bar{f}_o \bar{v}_t = q_2 \qquad (21)$$

(2) Part of the water in Infinitesimal A displaces that of Infinitesimal B, as shown in Fig.5 Part I, and the other part displaces oil in Infinitesimal B, as shown in Fig.5 Part III:

$$\bar{v}_w = \bar{f}_w \bar{v}_t = q_1 + q_3 \qquad (22)$$

(3) Inside the Infinitesimal B, whose oil is displaced by the oil in Infinitesimal A, and by part of the

water in Infinitesimal A, as shown in Fig. 5-Part II, III. Therefore,
$$v_o^+ = f_o^+ v_t = q_2 + q_3 \tag{23}$$

(4)All the water in Infinitesimal B only engages in in-phase displacement, as shown in Fig.5-Part I:
$$v_w^+ = f_w^+ v_t = q_1 \tag{24}$$

Because it is well known that $v_o + v_w = q_1 + q_2 + q_3$, the four equations of Equations (13)-(16) are not independent. The three variables q1, q2, q3 can be found through solving any three equations of Equations (13)-(16).

$$q_1 = f_w^+ v_t, \quad q_2 = (1 - f_w^-) v_t, \quad q_3 = (f_w^- - f_w^+) v_t \tag{25}$$

Substitute formula (4) into (10)-(13), obtains the interface condition for mass.

$$\begin{cases} v_t = v_o^- + v_w^- = v_o^+ + v_w^+ = v_t \\ v_o^- - v_o^+ = -\left[ f_w^- - f_w^+ \right] v_t \\ v_w^- - v_w^+ = \left[ f_w^- - f_w^+ \right] v_t \end{cases} \tag{26}$$

Clearly, Equation. (22) and (24) conform to the seepage differential equations of MUSKAT. In the same way, when oil displaces water, same result can also be reached.

## 3.2 Deducing the Interface Condition for Pressure

In the range of ($x_1, x_\Gamma$), via Formula (10), get the following:
$$P_1 = P^- + v_t \int_{x_1}^{x_\Gamma} \frac{1}{\lambda_t K} dx \tag{27}$$

From Formula (10), directly get the following:
$$P_2 = P_1 - v_t \left( \int_{x_1}^{x_\Gamma} \frac{1}{\lambda_t K} dx + \int_{x_\Gamma}^{x_2} \frac{1}{\lambda_t K} dx \right) \tag{28}$$

Substitute Formula (28) into (27), and get:
$$P_2 = P^- - v_t \int_{x_\Gamma}^{x_2} \frac{1}{\lambda_t K} dx \tag{29}$$

In the range of ($x_\Gamma, x_2$), from Formula (10), we get:
$$v_t = -\frac{P_2 - P^+}{\int_{x_\Gamma}^{x_2} \frac{1}{\lambda_t K} dx} \tag{30}$$

Namely:
$$P^+ = P_2 + v_t \int_{x_\Gamma}^{x_2} \frac{1}{\lambda_t K} dx \tag{31}$$

Substitute Formula (30) into (29) and get:
$$P^+ = P^- \tag{32}$$

Equation (32) is the interface conditions for pressure.

# 4. Application Examples of JVCM

Select a rock with a cross section area of 1m2 and a length of 100m from the formation, and place it horizontally. In the original state, the rock porosity contains 100% oil. Where x=0, the pressure is 20Mpa, joining in an end of the rock with infinite water body, with the pressure staying at 20Mpa, supplying water for the rock at a flux of 1m3/day, without regard to the influence of capillary force and gravity. The rock and fluid parameters are shown in Table 1. The relative permeability curve conforms to Corey relation(Brooks, R. H., & Corey, A. T. (1964). Hydraulic properties of porous media and their relation to drainage design. Transactions of the ASAE, 7(1), 26-0028.):

$$K_{rw}(S_{wn}) = S_{wn}^2, K_{row}(S_{wn}) = (1-S_{wn})^3 \tag{33}$$

where $S_{wn} = \dfrac{S_w}{1-S_{wi}-S_{orw}}$

The under-mentioned three seepage issues are solved by the use of analytic solution, displacement weighted jump condition model, conventional jump condition model and the standard single-point upstream weighted method (SPU).

Table.1 Fluid and physical properties in application examples.

| Parameters | Value |
|---|---|
| Relative permeability K | 300mD |
| porosity, ϕ | 0.15 |
| Oil viscosity, $\mu_o$ | 2mPas.s |
| Water viscosity, $\mu_w$ | 0.3mPa.S |
| cross section area A | 1m² |
| reservoir length | 100m |
| initial water saturation, Swi | 0 |
| Residual oil saturation, Sor | 0 |

## 4.1 Calculate the Front Flow

Assume that the position of the flood front interface is 71.41m and the pressure and saturation on the two points beside the flood front interface are listed in Table.2, and the relative permeability is referred to the relative permeability curve. The value flux as shown in Fig.3 can be calculated based on the known data.

The results have shown that (referred to Table 3, Fig.6)

(1) When the fluid is distributed continuously, calculation values by different methods all approach the accurate value;

(2) In flood front interface, the fluids present discontinuous distribution. Water flow and total flow predicted by employing the method in this paper (JVCM) is the closest to the accurate solution, with the errors of 0.68% and 0.35% respectively; whereas, the error in total flow deduced by the conventional Jump Condition (HRM) is 41.2% and that in water flow 0, which obviously makes no sense; the errors in total flow and water flow calculated through standard upstream weighted method are 23.14% and 24.41%.

Table 2 The pressure, saturation and relative permeability beside the flood front

| Location point | Coordinate | Saturation | Pressure | Krw | Krow |
|---|---|---|---|---|---|
| Inside point | 70.41316 | 0.182532 | 26.29014 | 0.023333 | 0.546313 |
| Outside point | 72.41316 | 0 | 25.51955 | 0 | 1 |

Table 3 Water flow and total flow calculated by different methods

| Location | Total flow | | | | Water flow | | | |
|---|---|---|---|---|---|---|---|---|
| | Accurate value | JVCM | SPU | HRM | Accurate value | JVCM | SPU | HRM |
| Inside point | 1 | 0.9831 | 0.9936 | 0.9829 | 0.6308 | 0.6264 | 0.6202 | 0.6199 |
| Front | 1 | **0.9932** | **1.2314** | **0.5881** | 0.6243 | **0.6265** | **0.7767** | **0.0000** |
| Outside point | 1 | 1 | 1 | 1 | 0 | 0 | 0 | 0 |

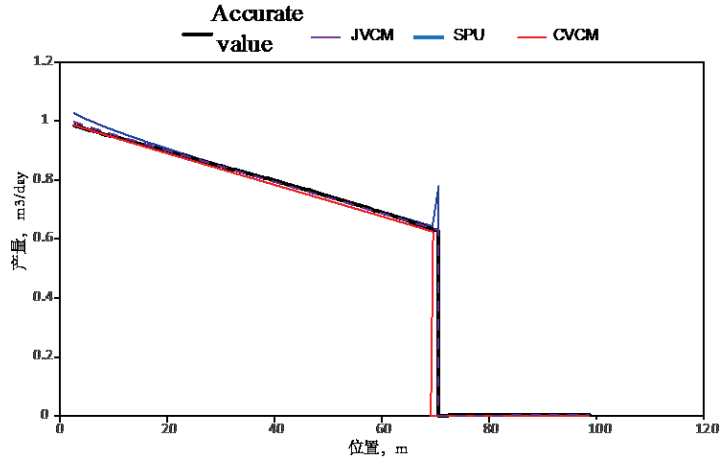

Fig.6 Flux distribution calculated by Different methods

### 4.2 Calculate the Pressure on the Flood Front

Suppose the flux is known as 1m3/day and the pressure of the outside point in Table 2 is unknown, and the results calculated by above-mentioned 4 methods are shown in Table 4 and summarized as follows:

(1) Pressure differences calculated by various methods are evidently differential, in which the result calculated by the method discussed in this paper (JVCM) approaches the accurate value most;

(2) Although the difference value calculated by different methods is large, the background pressure is still high and the pressure value difference is small. It is believed in this paper that via the comparison with different Jump Condition model, in general, pressure value is not adopted as comparing parameter.

Table.4 Pressure inside and outside of the flood front calculated by different methods (Flux: 1m3/day)

| Pressure of the point outside front, MPa | | | | Pressure Difference of the outside and inside points, MPa | | | |
|---|---|---|---|---|---|---|---|
| Accurate value | JVCM | SPU | HRM | Accurate value | JVCM | SPU | HRM |
| 25.9825 | 25.5143 | 25.6644 | 26.2246 | 0.770587 | 0.7759 | 0.6258 | 1.3104 |

### 4.3 Estimate Flood Front Position

Estimating front location is pivotal to the process of reservoir development, and the key to front tracer in reservoir numerical simulation(Hyman, 1984), especially to multiscale simulation(Chu, Engquist,

Prodanović, et.al. 2013; Jenny, Lee, Tchelepi, 2005). Supposing the flux is known as 1m3/day, while the accurate front location is unknown, it can be calculated by the date in Table 2, which ends up with different results by different methods, as shown in Table 5:

(1) The flood front interface predicted by the method mentioned in this paper is very close to the accurate solution; whereas predicted location by the conventional Jump Condition is not within the two points, which is obviously not reasonable.

(2) If the cross section area is equal to permeability, then the standard single-point upstream weighted (SPU) flux is irrelevant with the front interface. In other words, this condition cannot ascertain front interface.

Table.5 the Coordinate, pressure, saturation and relative permeability beside the flood front

| Point of location | Coordinate | Saturation | Pressure | Krw | Krow |
|---|---|---|---|---|---|
| Inside point | 70.41316 | 0.182532 | 26.29014 | 0.023333 | 0.546313 |
| Outside point | 80.41316 | 0 | 21.81585 | 0 | 1 |

Table.6 The position of the flood front predicted by different methods

| Location | Accurate value | JVCM | SPU | HRM |
|---|---|---|---|---|
| Relative location | 1.0000 | 1.0351 | - | -0.4040 |
| Coordinate location | 71.4100 | 71.4482 | - | 70.0090 |

## 5. Conclusions

(1) The mass conservation principle is not real foundation for the velocity conservation. That the Darcy velocity of each fluid on the both sides of jump interface is not equivalent does not implies the conditions for the Darcy velocities violate the mass conservation law; on the contrary, in light of the mass conservation principle, only when $v_w^- > v_w^+$, the oil within the region $[x_{\Gamma^-}, x_{\Gamma^+}]$ can be displaced by the water.

(2) The JVCM is more reasonable than CVCM. Moreover, in light of JVCM, it can be understood that why SPU causes big errors at discontinuous interface.(Tveit, Aavatsmark, 2012)

(3) The research result in this paper is easy to promote more general situations: multi-dimensional, or other types of discontinuous interface; if the continuous interface is taken as one special kind of discontinuous interface, JVCM can be used in any interface that has no source or sink term.

In addition, in the conventional Jump Condition, the equivalent relation of each phase fluid pressure besides the flood front is problematic. Due to space limitation, the author will discuss this in the following paper.

## Acknowledgements

This work is supported by the National Science and Technology Major Project of the Ministry of Science and Technology of China2011ZX05060-004 and 2011ZX0514-004.